\documentclass[preprint,prd]{revtex4}
\usepackage{epsfig}
 \usepackage{axodraw}

\begin{document}

\title{Direct CP Violation in Hadronic B Decays
\footnote{work supported by national science foundation of China.}
}
\author{Bi-Hai Hong}
\affiliation{Department of Physics, Lishui University, Lishui,
Zhejiang 323000, China; \\
 Department of Physics, Shanghai Jiaotong
University, Shanghai 200030, China}

\author{  Cai-Dian L\"u}

\affiliation{Institute of High Energy Physics,   P.O. Box 918(4),
Beijing 100049, China}

\begin{abstract}
 There are different approaches for the hadronic B decay calculations, recently.
 In this paper, we upgrade three of them, namely
factorization, QCD factorization and the perturbative QCD approach
 based on $k_T$ factorization, by using new parameters and full wave functions.
 Although they get similar results for many of the branching ratios,
 the direct CP asymmetries predicted by them
 are different, which can be tested by recent experimental measurements of B factories.

\end{abstract}


\maketitle

\section{Introduction}

The hadronic B decays are important for testing the standard model
(SM), and also for uncovering the signal of new physics.
Understanding non-leptonic $B$ meson decays is crucial for  the
CKM matrix elements measurements and CP violation detection.
Especially the direct CP violation is expected to be measured in
hadronic B decays. Recently, both Belle and BaBar claimed to find
direct CP asymmetry in $B^0\to \pi^+\pi^-$ and $B^0 \to \pi^+ K^-$
decays \cite{cpa}. It is the start of direct CP measurement in B
physics.

In the theoretical side, there are much more complication in
direct CP asymmetry predictions. The direct CP violation in
hadronic decays requires at least two
 decay amplitudes with different weak phase and strong phase.
In the standard model, the weak phase comes from the so called CKM
matrix, which is well defined.  Although the weak phase from SM is
clean, the strong phase requires hadronic matrix element
calculation, which is usually model dependent.

The simplest case is two-body hadronic $B$ meson decays, for which
Bauer, Stech and Wirbel (BSW) proposed the naive factorization
assumption (FA) in their pioneering work \cite{BSW}. Considerable
progress, including generalized FA \cite{akl,Cheng94} and
QCD-improved FA (QCDF) \cite{BBNS}, has been made. On the other
hand, technique to analyze hard exclusive hadronic scattering was
developed by Brodsky and Lepage \cite{LB} based on collinear
factorization theorem in perturbative QCD (PQCD). A modified
framework based on $k_T$ factorization theorem was then given in
\cite{BS,LS}, and extended to exclusive $B$ meson decays in
\cite{LY1,li,CLY}.

The predictions of branching ratios agree well with experiments in
most cases,   thus it is difficult to tell from experiments that
which method is better than others. However, the strong phase,
which is important for the CP violation prediction, is quite
sensitive to
 various approaches. The mechanism of this strong phase is quite
 different for various method, and give quite different results.
The recent experimental results \cite{cpa} can make a test for the
validity of these approaches.

In this paper, we will first introduce the factorization approach
in next section. The QCD factorization and improved PQCD approach
based on $k_T$ factorization are then introduced in section
\ref{sec3} and section \ref{sec4}, respectively. In section
\ref{sec5}, we upgrade the numerical results of these approaches
using newest parameters. We  compare the three major approaches to
show the difference of direct CP asymmetry predicted by them.
Finally the summary is presented.

\section{Naive and generalized factorization approach}

 The calculation of
non-leptonic decays involves the short-distance Lagrangian and the
calculation of hadronic matrix elements which are model dependent.
 The short-distance QCD corrected Lagrangian is  calculated
 to  next-to-leading order \cite{buras}.
 In 1987, Bauer, Stech and Wirbel first calculate the hadronic $B$
and $D$ decays using the naive factorization approach \cite{BSW}.
In their factorization method, the hadronic matrix element is
expressed as a product of two factors
 $\langle h_1h_2 | {\cal H}_{eff}|B\rangle =\langle h_1 | J_1|B \rangle
\langle h_2 | J_2|0 \rangle $.    The first factor is proportional
to the $B\to h_1$ form factor, while the second one is
proportional to the decay constant of $h_2$ meson. All the
perturbative channel dependent part is described by the Wilson
coefficients of four quark operators.

  These effective four-quark operators are induced by the weak interaction in the quark level, which
  gives  the short distance contribution to the non-leptonic
  decays of B mesons.
The effective Hamiltonian for the charmless  non-leptonic B decays
is \cite{buras}
\begin{eqnarray}
\label{heff} {\cal H}_{eff} = \frac{G_{F}}{\sqrt{2}} \, \left[
V_{q'b} V_{q'q}^* \, \left(\sum_{i=1}^{10} C_{i} \, O_i + C_g O_g
\right) \right]  ,
\end{eqnarray}
where $q=d,s$ and $V_{q'q}$ denotes the CKM factors. The operators
$O_1,O_2$ are tree level current operators. The operators
$O_3,\ldots,O_6$ are QCD penguin operators. $O_7,\ldots,O_{10}$
arise from electroweak penguin diagrams, which are suppressed by
$\alpha/\alpha_s$. Only $ O_9$ has a sizable value whose
 major contribution arises from the $Z$ penguin. The $C_i$'s are
 the Wilson coefficients of four quark operators with  QCD
  corrections.

Although this is a very simple method, later experiments show that
many of the decay branching ratios explained well by the FA
\cite{neubert}, especially for the color enhanced decays class I,
III and class IV \cite{akl}. When charmless $B$ meson decays are
considered, efforts have been made to generalize the FA approach
\cite{akl,Cheng94}.  To explain the non-factorizable dominated
class II and class V decays, phenomenological parameter
$N_c^{eff}$ is introduced. Most of the branching ratios of
hadronic $B$ decays agree well with experiments by $N_c^{eff}=2$
\cite{akl,Cheng94}.

An effort is also made to predict the CP violation parameters in
different hadronic $B$ decays \cite{akl2}. Since direct CP
violation requires a strong phase difference between decay
amplitudes in addition to a weak phase difference, the precision
of strong phase calculation is essential. In naive FA, the
hadronization is described by the form factor only, therefore, no
strong phase is given.

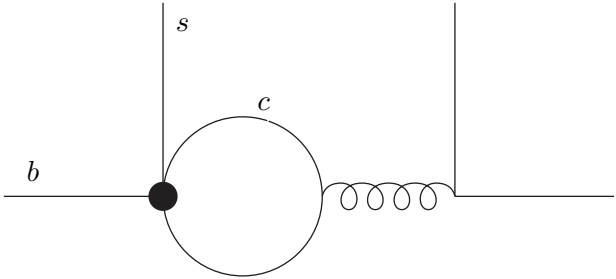
\begin{figure}[tbp]
   \begin{picture}(130,160)(0,20)
        \Line(60,72)(0,72)
        \Line(60,72)(60,145)
        \CArc(90,72)(30,72,71)

        \Gluon(120,72)(170,72){5}{4} \Vertex(60,72){5.5}
        \Line(170,72)(230,72)
        \Line(170,72)(170,145)

       \put(65,135){$s$}
      \put(96,105){$c$}
        \put(5,78){ {${b}$}}
   \end{picture}
 \caption{The perturbative charm quark loop diagram generating
 strong phases in FA and QCDF.}
\label{fig1}
\end{figure}

In generalized FA \cite{akl2}, the next-to-leading order Wilson
coefficients contain strong phases generated by the charm quark
loop, where charm quark can be on shell. The diagram shown in
Fig.\ref{fig1} is also called BSS mechanism \cite{bss}. The size
of this kind of strong phase is sensitive to the momentum of the
gluon connecting to the charm quark loop. However, in FA, this
momentum is not well defined, since all hadronic dynamics are
defined only by form factors. In ref. \cite{akl2}, the authors use
$k^2 = m_b^2/2 \pm 2 $GeV$^2$ to give the CP asymmetry parameters
of many channels.

\section{QCD Factorization Approach}
\label{sec3}

\begin{figure}[t]
\epsfig{file=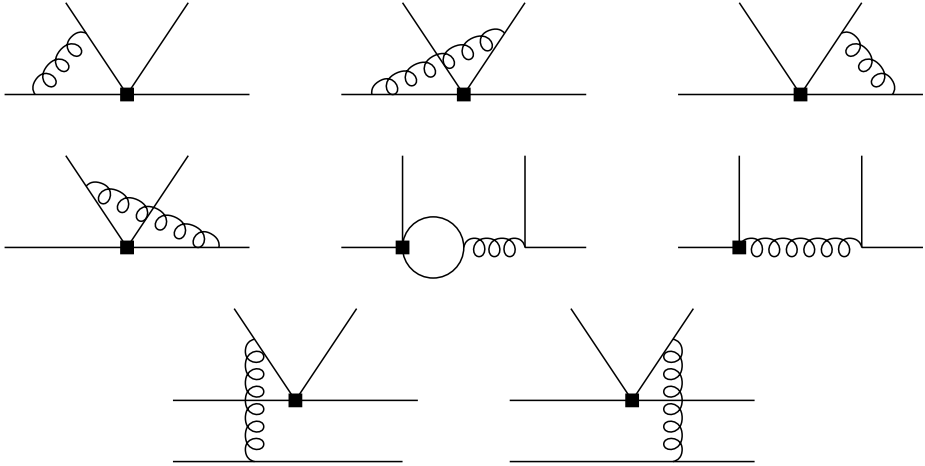,bbllx=5cm,bblly=10cm,bburx=15.5cm,bbury=25cm,%
width=7cm,angle=270} \caption{Figures calculated in the QCD
factorization approach. \label{figbbns}}
\end{figure}

In 1999, Beneke, Buchalla, Neubert, and Sachrajda (BBNS) proposed
a formalism for two-body charmless $B$ meson decays \cite{BBNS}.
In this approach, they expand the hadronic matrix element by the
heavy b quark mass and $\alpha_s$
\begin{equation}
\langle \pi \pi | O_i | B \rangle = \langle \pi |j_1 |B\rangle
\langle \pi |j_2| 0\rangle \left[ 1+ \sum r_n \alpha_s^n + {\cal
O} (\Lambda_{QCD} /m_b) \right].\label{bbnsf}
 \end{equation}
 Here $O_i$ is a local
operator in the effective Hamiltonian and $j_{1,2}$ are bilinear
quark currents. By neglecting the power corrections of ${\cal O}
(\Lambda_{QCD})$, one need only calculate the order $\alpha_s$
corrections including the vertex corrections for the four quark
operators and the non-factorizable diagrams. These diagrams are
shown in Fig.\ref{figbbns}. The first 6 diagrams have already been
included in the generalized FA approach as next-to-leading order
QCD corrections to local four quark operators. What new are the
last two non-factorizable diagrams, which has a hard gluon line
connecting the four quark operator and the spectator quark.

They claimed that factorizable contributions, for example, the
form factor $F^{B\pi}$ in the $B\to\pi\pi$ decays, are dominated
by long distance contributions.  Hence, it is treated in the same
way as FA, and expressed as products of Wilson coefficients and
form factor $F^{B\pi}$. The   non-factorizable contributions
calculated perturbatively, are written as the convolutions of hard
amplitudes with three $(B,\pi,\pi)$ meson wave functions.
Annihilation diagrams are neglected as in FA.  Values of form
factors at maximal recoil $q^2=m_\pi^2$ and non-perturbative meson
wave functions are all treated as input parameters. It is easy to
see from eq.(\ref{bbnsf}), that this equation is only applicable
for those color enhanced decay modes, where the factorizable
contribution dominates the final results. While for the color
suppressed modes, where the non-factorizable contributions   are
not small, the expansion of eq.(\ref{bbnsf}) is not right, since
the large non-factorizable contribution is grouped into the
next-to leading order term of eq.(\ref{bbnsf}).

  The numerical results show that the
theory and experiments agree well for those class I and IV decays,
which are color enhanced and dominated by the factorizable
contribution. This also agrees with the FA result, since the
dominant part in eq.(\ref{bbnsf}) is the same as the FA. The
success of QCD factorization is that one can calculate the
sub-leading ${\cal O} (\alpha_s)$ non-factorizable contribution
(second term in eq.(\ref{bbnsf})) using perturbative QCD. While in
the FA, one has to input a free parameter $N_c^{eff}$ to
accommodate the non-factorizable contribution.

In the QCD factorization calculations,   people found that there
exist endpoint divergence in the annihilation diagram calculations
  \cite{suda}. Logarithm divergence occurred at
twist 2 contribution, and linear divergence exists in twist 3
contribution. If not symmetric wave function, like $K^{(*)}$
meson, there is also soft divergence in the non-factorizable
diagrams. It is very difficult to treat these singularity in the
BBNS approach. A cut-off is introduced to regulate the divergence,
thus makes the QCD factorization approach prediction parameter
dependent especially in the annihilation diagrams.

As for the strong phase in BBNS, like in FA, it mainly comes from
the BSS mechanism. Here the momentum of the inner gluon is well
defined, since wave functions are introduced. However, it predicts
too small strong phase, because of the small gluon momentum.
Hence, small direct CP asymmetry is predicted. There is also
another source of strong phase from the annihilation diagrams, but
strongly depends on the cut-off parameter. The strong phase  in
QCD factorization can be
    large due to this cut-off.

In a word, the QCD factorization approach is at least one step
forward from Naive Factorization approach. It gives systematic
prediction of sub-leading non-factorizable contribution for the
class I and class IV decays, which are dominated by the
factorizable contribution.   Problem remained is the endpoint
singularity in higher order calculations, but may be solved with
Sudakov form factors like PQCD approach.

\section{Formalism of PQCD Approach}
\label{sec4}

 In this section, we will introduce the idea of PQCD
approach. The three scale PQCD factorization theorem has been
developed for non-leptonic heavy meson decays \cite{li}, based on
the formalism by Brodsky and Lepage \cite{LB}, and Botts and
Sterman \cite{BS}. In the non-leptonic two body B
 decays, the $B$ meson is heavy, sitting at rest.
It decays into two light mesons with large momenta. There must be
a hard gluon to kick the light spectator quark (with small
momentum) in the B meson to form a fast moving light
 meson. So the dominant diagram in this theoretical picture
is that one hard gluon from the spectator quark connecting with
the other quarks in the four quark operator of the weak
interaction. Unlike the usual factorization approach, the hard
part of the PQCD approach consists of six quarks rather than four.
We thus call it six-quark operators or six-quark effective theory.
There are also infrared (soft and collinear) gluon exchanges
between quarks. Summing over those leading soft contributions
gives a Sudakov form factor, which suppresses the soft
contribution. Therefore, it makes the PQCD reliable in calculating
the non-leptonic decays. With the Sudakov resummation, we can
include the leading double logarithms for all loop diagrams, in
association with the infrared contribution.

There are three different scales in the B meson non-leptonic
decays $M_W$, $m_b$ and $1/b$. The first scale describe the
intrinsic electroweak decay of $B$ meson, through charged current.
The second scale $m_b$ denote the scale of energy release in the
decay. Since the $b$ quark decay scale $m_b$ is much smaller than
the electroweak scale $m_W$, the QCD corrections to the four quark
operators are non-negligible, which are usually summed by the
renormalization group equation. This has already been done to the
leading logarithm and next-to-leading order for years
\cite{buras}. The third scale $1/b$ involved in the $B$ meson
exclusive decays is usually called the factorization scale, with
$b$ the conjugate variable of parton transverse momenta. The
dynamics below $1/b$
 scale is regarded as being completely
non-perturbative, and can be parameterized into meson wave
functions. The meson wave functions are not calculable in PQCD.
But they are universal, channel independent. We can determine them
 from experiments, and they are constrained   by QCD sum rules and
Lattice QCD calculations. Above the scale $1/b$, the physics is
channel dependent. We can use perturbation theory to calculate
channel by channel.

With all the large logarithms resummed, the remaining finite
contributions are absorbed into a perturbative b quark decay
sub-amplitude $H(t)$. Therefore the three scale factorization
formula is given by the typical expression,
\begin{equation}
C(t) \times H(t) \times \Phi (x) \times \exp\left[ -s(P,b) -2 \int
_{1/b}^t \frac{ d \bar\mu}{\bar \mu} \gamma_q (\alpha_s (\bar
\mu)) \right], \label{eq:factorization_formula}
\end{equation}
where $C(t)$ are the corresponding Wilson coefficients, $\Phi (x)$
are the  meson wave functions and the variable $t$ denotes the
largest mass scale of hard process $H$, that is, six-quark
effective theory.
 The quark anomalous dimension $\gamma_q=-\alpha_s /\pi$ describes
the evolution from scale $t$ to $1/b$. Since logarithm corrections
have been summed by renormalization group equations, the above
factorization formula does not depend on the renormalization scale
$\mu$ explicitly.

As shown above, in the PQCD approach, we keep the $k_T$ dependence
of the wave function. In fact, the approximation of neglecting the
transverse momentum can only be done at the non-endpoint region,
since $k_T \ll k^+$ is qualified  at that region. At the endpoint,
$k^+ \to 0$, $k_T$ is not small any longer, neglecting $k_T$ is a
very bad approximation. By, keeping the $k_T$ dependence,
 there is no endpoint divergence as
occurred in the QCD factorization approach, while the numerical
result does not change at other region. Furthermore, the Sudakov
form factors suppress the endpoint region of the wave functions.

        \begin{figure}[tbp]
\epsfig{file=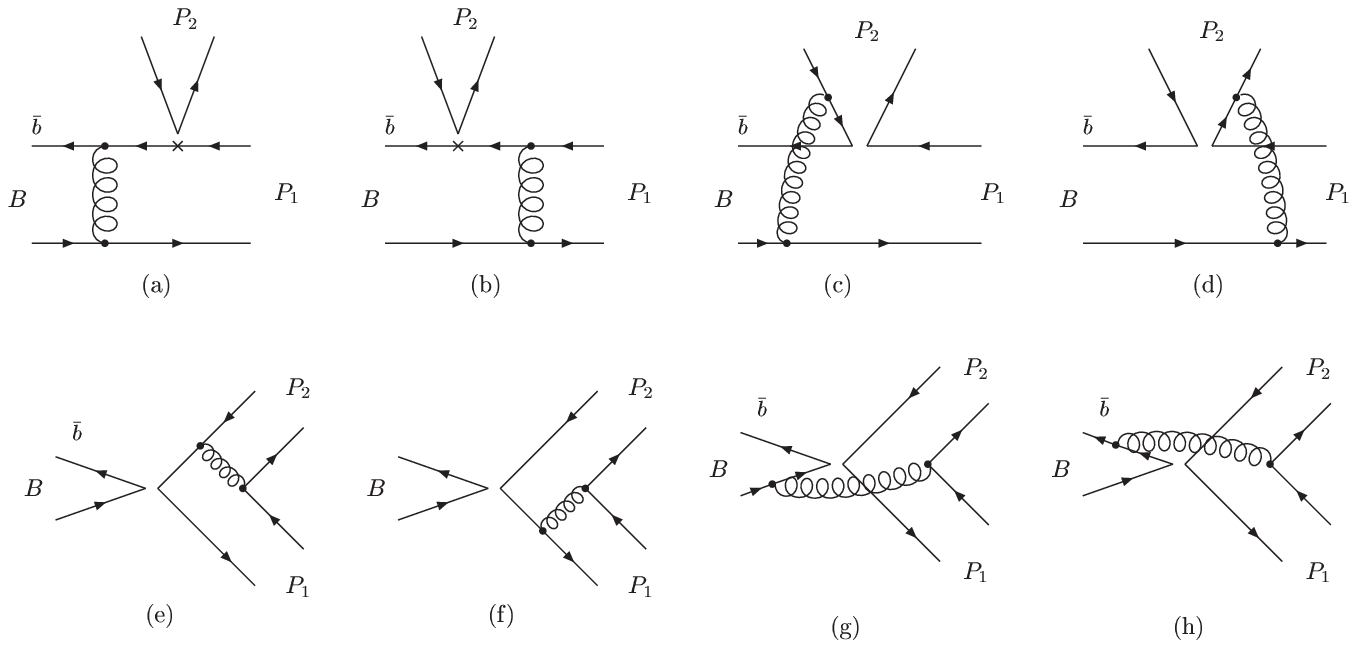,bbllx=5.6cm,bblly=18.9cm,bburx=16.5cm,bbury=25cm,%
width=10.8cm,angle=0}
    \caption{Diagrams for $B\to P_1P_2$ decay in perturbative QCD approach.
     The factorizable
    diagrams (a),(b), non-factorizable (c),
    (d),  factorizable annihilation
    diagrams (e),(f) and non-factorizable annihilation diagrams (g),(h).}
    \label{fig2}
   \end{figure}

The main input parameters in PQCD are the meson wave functions. It
is not a surprise that the final results are   sensitive to the
meson wave functions. Fortunately, there are many channels involve
the same meson, and the meson wave functions should be process
independent. In all the calculations of PQCD approach, we follow
the rule, and we find that   they can explain most of the measured
branching ratios of B decays.   For example: $B\to \pi\pi$ decays,
$B\to \pi \rho$,
  $B\to \pi \omega$ decays \cite{LUY}, $B\to K\pi$ decays \cite{KLS},
   $B\to K K$ decays
  \cite{bkk}, the form factor calculations of $B\to \pi$, $B\to \rho$
  \cite{semi}, $B\to K \eta^{(\prime)}$ decays \cite{eta},
   $B\to K\phi$ decays \cite{bkphi}    etc.

We emphasize that non-factorizable (Fig.\ref{fig2}(c)(d)) and
annihilation diagrams (Fig.\ref{fig2}(e-h)) are indeed sub-leading
in the PQCD formalism as $M_B\to \infty$. This can be easily
observed from the hard functions in appendices of
ref.\cite{KLS,LUY}. When $M_B$ increases, the $B$ meson wave
function enhances contributions to factorizable diagrams. However,
annihilation amplitudes, being independent of B meson wave
function, are relatively insensitive to the variation of $M_B$.
Hence, factorizable contributions become dominant and annihilation
contributions are sub-leading in the $M_B\to\infty$ limit
\cite{bkphi}. Although the non-factorizable and annihilation
diagrams are sub-leading for the branching ratio in color enhanced
decays, they provide the main source of strong phase, by inner
quark or gluon on mass shell. The BSS mechanism should also be
present   in the PQCD approach. However this mechanism of strong
phase is small, since it is at next-to-leading order $O(\alpha_s)$
corrections.

\section{Numerical Results and Discussion}
\label{sec5}

In numerical analysis, we use recently updated decay constants and
form factors for the FA and QCDF approaches. The branching ratios
changes a little bit, since they are sensitive to these
parameters. But the direct CP asymmetry changes very little. In
the PQCD approach, we use the full set of light meson wave
functions, including two twist 3 distribution amplitudes, where
only one used in the previous papers \cite{LUY,KLS}. The threshold
resummation for the endpoint of the hard part calculation is also
newly included \cite{thr}. The numerical results for the branching
ratios and CP asymmetries change only a little bit.

\begin{table}[t] \caption{Direct CP asymmetries calculated in
FA \cite{akl2}, QCDF \cite{BBNS} and PQCD \cite{LUY,KLS} for $B\to
\pi\pi$ and $B\to K\pi$ decays together with the averaged
experimental results at percentage.}{\begin{tabular}{|l| c| c|c|c|
}
\hline Quantities & FA & QCDF & PQCD & Data \\
\hline
$ B^0\to \pi^+\pi^-$ &      $-5\pm 3$   &    $-6\pm 12$ &
 $+30\pm 20$ &    $+37\pm 11$ \\
$ B^0\to \pi^+K^-$ &$+10\pm 3$  &$+5\pm 9$  & $-17\pm 5$ &$-10.9\pm 1.9$\\
$B^+\to K^0\pi^+$ & $+1.7\pm 0.1$  &$ +1 \pm 1$ &$  -1.0\pm 0.5$
  &$-2.0 \pm 3.4$ \\
$B^+\to K^+\pi^0$ & $ +8\pm 2$  &$+ 7 \pm 9$ &$  -13\pm 4$   &$ +4 \pm 4$ \\
\hline
\end{tabular}}
\label{cp}
\end{table}

As discussed in the previous section, the strong phase generated
 from PQCD approach is quite different from the FA and QCDF
approaches. The direct CP asymmetry in SM is proportional to the
sine of the strong phase difference of two amplitudes. Therefore
the direct CP asymmetry will be different if strong phase is
different. The predicted CP asymmetry by the three methods are
shown in table \ref{cp}. It is easy to see that the FA \cite{akl2}
and QCDF \cite{BBNS} results are quite close to each other, since
the mechanism of strong phase is the same for them.

Recently the two B factories measure some   channels with non-zero
direct CP asymmetry \cite{cpa}, which are shown in table \ref{cp}
\cite{keum}. It is claimed that direct CP has been found in
$B^0\to \pi^+\pi^-$ and $B^0\to \pi^+K^-$ decays with more than
$4\sigma$ signal. It is easy to see that our PQCD results of
direct CP asymmetry \cite{LUY} agree with the experiments,
especially for the experimentally well measured channels $ B^0\to
\pi^+\pi^-$ and $ B^0\to \pi^+K^-$ decays. Although FA and QCDF
are not yet ruled out by experiments, but the experiments at least
tell us that the dominant strong phase should come from the
mechanism of PQCD not QCDF. Charm quark loop mechanism, which
gives the central value of strong phase in QCDF, is argued next-to
leading order in PQCD. This argument is now proved by B factories
experiments.

In summary, branching ratios of both FA and QCDF depend on the
values of form factors. The strong phase generated from these
approaches is dominantly from perturbative BSS mechanism.
 In the PQCD approach, dominant by short distance contribution, the form factors are
sensitive to meson wave functions. By including the $k_T$
dependence and Sudakov suppression, there is no endpoint
divergence. In the PQCD formalism, annihilation  amplitudes are of
the same order as non-factorizable ones in powers of $1/M_B$,
which are both $O(1/(M_B\Lambda_{\rm QCD}))$. The strong phase
comes mainly from the annihilation and non-factorizable diagrams
in PQCD approach, which is quite different from the FA and QCDF
approaches. The experimentally measured direct CP asymmetry
implies that PQCD gives at least the dominant strong phase than
other approaches. This will be further tested by experiments.

\end{document}